\documentclass[reprint,superscriptaddress,amsmath,amssymb,aps,pra]{revtex4-1} 
\pdfoutput=1

\usepackage{graphicx}
\usepackage{dcolumn}
\usepackage{bm}
\usepackage{braket}
\usepackage{color}
\usepackage{hyperref}

\usepackage[hang,small,bf]{caption}
\usepackage[subrefformat=parens]{subcaption}
\makeatletter
\newcommand{\printfnsymbol}[1]{%
  \textsuperscript{\@fnsymbol{#1}}%
}
\makeatother

\begin{document}

 \title{Solvent distribution effects on quantum chemical calculations with quantum computers}

\author{Yuichiro Yoshida }
\thanks{Y.Y. and W.M. contributed equally.}
\email{yoshida.yuichiro.qiqb@osaka-u.ac.jp}
\affiliation{Center for Quantum Information and Quantum Biology, Osaka University, 1-2 Machikaneyama, Toyonaka, Osaka, 560-0043, Japan}

\author{Wataru Mizukami \printfnsymbol{1}}
\email{wataru.mizukami.857@qiqb.osaka-u.ac.jp \textcolor{black}{(corresponding author)}}
\affiliation{Center for Quantum Information and Quantum Biology, Osaka University, 1-2 Machikaneyama, Toyonaka, Osaka, 560-0043, Japan}
\affiliation{Graduate School of Engineering Science, Osaka University, 1-3 Machikaneyama, Toyonaka, Osaka 560-8531, Japan.}

\author{Norio Yoshida}
\email{noriwo@nagoya-u.jp}
\affiliation{Department of Chemistry, Graduate School of Science, Kyushu University, 744 Motooka, Nishiku, Fukuoka 819-0395, Japan.}
\affiliation{Department of Complex Systems Science, Graduate School of Informatics, Nagoya University, Furo-cho, Chikusa-ward, Nagoya 464-8601, Japan.}

\date{\today}

\begin{abstract}
We present a combination of three-dimensional reference interaction site model self-consistent field (3D-RISM-SCF) theory and the variational quantum eigensolver (VQE) to consider the solvent distribution effects within the framework of quantum-classical hybrid computing. 
The present method, 3D-RISM-VQE, does not include any statistical errors from the solvent configuration sampling owing to the analytical treatment of the statistical solvent distribution.
We apply 3D-RISM-VQE to compute the spatial distribution functions of solvent water around a water molecule, the potential and Helmholtz energy curves of NaCl, and to conduct Helmholtz energy component analysis of H$_2$O and NH$_4^+$.
Moreover, we utilize 3D-RISM-VQE to analyze the extent to which solvent effects alter the efficiency of quantum calculations compared with calculations in the gas phase using the $L^1$-norms of molecular electronic Hamiltonians.
Our results demonstrate that the efficiency of quantum chemical calculations on a quantum computer in solution is virtually the same as in the gas phase.
\end{abstract}

\maketitle

\section{\label{sec:intro} Introduction}

Quantum computers have the potential to reveal the electronic structure of molecules that are intractable using conventional methods\cite{Reiher2017elucidating}.
Quantum computers are classified into fault-tolerant quantum computers and noisy intermediate-scale quantum (NISQ) devices\cite{Preskill2018quantumcomputingin}.
Fault-tolerant quantum computers require a tremendous amount of qubits to enable error-correction schemes that protect qubits from noise and, thus, take a very long time to develop.
Quantum phase estimation (QPE) is a well-known quantum algorithm for fault-tolerant quantum computers.
On the other hand, NISQ devices are quantum devices with a few dozen to a hundred or more qubits and do not feature error-correction schemes.
Quantum-classical hybrid algorithms combining a NISQ device and a conventional computer have been actively studied. 
The variational quantum eigensolver (VQE)\cite{peruzzo2014variational} is a representative and widely-used example of such an algorithm.
Pioneering research has used noisy quantum devices for quantum chemical energy calculations of small molecules in gas phase, e.g., HeH$^+$ with a photonic quantum device\cite{peruzzo2014variational}, H$_2$, LiH, and BeH$_2$ with a superconducting quantum processor\cite{kandala2017hardware}, and H$_2$O with a trapped-ion quantum computer\cite{nam2020ground}.

Since most chemical processes progress in solution, 
it is highly desirable to extend quantum chemical methods for quantum computers to include solvent effects.
Algorithm development on solvent environment effects in quantum computing is a nascent research direction but is expected to greatly expand the applicability of quantum computers in the future.
Thus far, studies have considered solvent effects within the framework of quantum computing. 
Castaldo et al. unified the VQE and an implicit solvent model, i.e., the polarizable continuum model (PCM)
~\cite{castaldo2021quantum}.
The PCM model has also been used by Blunt et al. to estimate the computational costs of the QPE for a protein--ligand binding system~\cite{blunt2022perspective}. 
Izsak et al. introduced a simple multilayer approach (the ONIOM approach, also known as subtractive QM/MM) for the VQE and QPE~\cite{izsak2022quantum}.
Parrish et al. used explicit solvent models such as the standard additive QM/MM approach ~\cite{parrish2021analytical,hohenstein2022efficient}.
However, in these studies, there is little discussion of the impact of solvation effects on the computational efficiency of quantum computation, although Izsak et al. and Blunt et al. have provided cost estimates for the QPE in the presence of solvents.

Computational cost is a core issue in the study of quantum algorithms.
Algorithm development for quantum chemical calculations using quantum computers has advanced rapidly in recent years because of the prospect that quantum computing can accelerate molecular electronic structure calculations.
This potential acceleration has been driving the development of quantum algorithms, necessitating the study of computational efficiency.
In this context, 
it is crucial to address the following question: 
to what extent does the efficiency of quantum computing change when solvent effects are taken into account?

A fundamental quantity for measuring quantum computational cost is the $L^1$-norm of the Hamiltonian in qubit representation.
The efficiencies of both the VQE and QPE largely depend on the $L^1$-norm~\cite{koridon2021orbital}.
In the case of fault-tolerant quantum computing, the gate complexities of non-Clifford gates, such as T and Toffoli gates, determine the wall-time.
For instance, the gate count of the QPE based on the linear combination of unitaries directly depends on the $L^1$-norm~\cite{berry2019qubitization}.
In the qDRIFT algorithm for Hamiltonian simulation utilizing a random compiler, the gate probabilities depend on the $L^1$-norm and its components\cite{campbell2019random}.
\textcolor{black}{Techniques for reducing $L^1$-norm value using symmetries or the interaction picture have been considered by Loaiza et al\cite{loaiza2022reducing}.}
In the framework of the VQE, the variance in the expectation value of the Hamiltonian (i.e., its energy) directly relates to the $L^1$-norm. 
Several \textcolor{black}{shot-optimization} techniques have been developed to determine the required number of measurements of the Hamiltonian in order to minimize the variance~\cite{wecker2015progress,Rubin2018,arrasmith2020operator}.
A summary of the $L^1$-norm dependence of the quantum algorithms can be 
found in Koridon et al.~\cite{koridon2021orbital}, whose study also shows that orbital localization can reduce the value of the $L^1$-norm.

How the cost of quantum computing varies with or without solvent molecules can be answered by examining differences in the Hamiltonians of liquid and gas phases.
Although the Hamiltonian of a molecule in the gas phase can be uniquely prepared,
it has large arbitrariness in the liquid phase.
Countless numbers of solvent molecules in solution can influence the electronic structure of a solute molecule.
Solvent molecules move as a result of complex solute--solvent and solvent--solvent interactions.
The properties of solute molecules in a solution are evaluated by the ensemble average of the configuration of solvent molecules.
It is crucial but challenging to evaluate solvent configurations because sampling configurations includes a statistical error.

An approach to avoid the arbitrariness of the electronic Hamiltonian in the liquid phase involves utilizing a statistical theory of solvation~\cite{ten1993hybrid,ten1994reference,sato1996analytical}.
This approach, initiated by Ten-no et al., is called reference interaction site model self-consistent field (RISM-SCF) theory. 
Since its development, RISM-SCF has been extended to incorporate a three-dimensional distribution function using the three-dimensional reference interaction site model (3D-RISM) theory~\cite{hirata2003molecular,beglov1997integral,kovalenko1998three}. 
The resulting theory is thus called the 3D-RISM-SCF theory~\cite{kovalenko1999self,sato2000self,okamoto2019implementation}.
These theories (RISM-SCF/3D-RISM-SCF) solve the Schr\"{o}dinger equation of solute electronic wave function and the RISM/3D-RISM integral equation and determine the electronic structure of a solute and the spatial distribution of solvents simultaneously in a self-consistent manner.
\textcolor{black}{These theories have been applied to various chemical processes, such as chemical reactions in solution, liquid--solid interfaces, biological processes, etc\cite{Yoshida2021}.}

In this paper, we present a combination method of VQE and 3D-RISM-SCF, termed 3D-RISM-VQE, and propose quantum computing of molecules surrounded by a statistically distributed solvent.
Because the 3D-RISM-VQE is free from the statistical error of the solvents' configuration,
it is considered a favorable method by which to assess solvent distribution effects on the VQE using a quantum processor.
We emulated small quantum devices using computers and investigated the solvent distribution effect on several systems by 3D-RISM-VQE.
The spatial distribution of water around H$_2$O and the potential and Helmholtz energy curves of NaCl are examined.
Then, the Helmholtz energy components of H$_2$O and NH$_4^+$ were analyzed.
Another contribution involves evaluating the quantum computational cost of the Hamiltonian modification to include the solvent effect.
$L^1$-norm analysis was performed to reveal the additional quantum computational costs required to include the solvent effect.

\section{\label{sec:theory} Theory and computational method}

\subsection{A brief review of 3D-RISM-SCF theory}

\subsubsection{Solvated Hamiltonian}

The solvated Hamiltonian $\mathcal{H}_{\rm solv}$ can be given by the Hamiltonian of the isolated solute $\mathcal{H}_{\rm iso}$ and the solute--solvent interaction term $\mathcal{V}_{\rm solv}$ using a point-charge model as follows:
\begin{align}
\mathcal{H}_{\rm solv} &:= \mathcal{H}_{\rm iso} + \mathcal{V}_{\rm solv} \\
&\,= \mathcal{H}_{\rm iso} - \sum_{i=1}^{N_e} \sum_{I=1}^{N_{\rm ref}} \frac{q_I}{|{\bm r_i}-{\bm r_I}|}, \label{eq:Hsolv}
\end{align}
where $N_e$ and $N_{\rm ref}$ are the number of electrons and reference points to represent the solvent structure, respectively.
${\bm r_i}$ is the coordinate of the $i$-th electron, ${\bm r_I}$ is that of the $I$-th reference point and $q_I$ is the solvent charge density represented as the point charge on the $I$-th reference point.
The solvent charge density $q_I$ contains the information on the spatial distribution of a solvent around a solute, or $g_{v}({\bm r_I})$, where $v$ is the index of the interactive atomic site of the solvent.
The distribution function $g_{v}({\bm r_I})$ is calculated by solving the 3D-RISM equation~\cite{hirata2003molecular}.

\subsubsection{3D-RISM equation}

A formulation of the 3D-RISM equation is as follows:
\begin{align}
g_v({\bm r}) = 1 + \sum_{v' \in {\rm solvent}} c_{v'}({\bm r'}) * \chi_{v' v}(|{\bm r'}-{\bm r}|), \label{eq:RISM}
\end{align}
where $c_{v}({\bm r})$ and $\chi_{v'v}$ are the direct correlation function of the solvent and the solvent susceptibility function, respectively.
The symbol $*$ denotes a convolution integral.
Hereafter, we may omit the index $I$ and represent ${\bm r_I}$ as ${\bm r}$ when obvious.
The 3D-RISM equation itself cannot be solved and requires another equation to provide the approximate relation between $c_{v}({\bm r})$ and $g_v({\bm r})$.
Such equations are termed closure relations; the hyper-netted chain and the Kovalenko--Hirata (KH) closures are representative of these relations.
KH closure\cite{kovalenko1999self} is given by:
\begin{align}
    g_v({\bm r}) &= 
    \begin{cases}
    \begin{matrix}
{\rm exp}(d_v({\bm r})) & (d_v({\bm r}) \leq 0), \\
1 + d_v({\bm r}) & (d_v({\bm r}) > 0), \label{eq:KH}
\end{matrix}
\end{cases}\\
d_v({\bm r}) &= -\beta u_v({\bm r}) + g_v({\bm r}) - c_v({\bm r}) - 1
\end{align}
where $\beta$ is an inverse temperature and $u_v({\bm r})$ is the interaction potential between the solute and solvent site $v$ at position ${\bm r}$.
The interaction potential is usually given by the sum of the Lennard-Jones and electrostatic terms as follows:
\begin{align}
    u_v({\bm r})&= \sum_{u \in {\rm solute}} 4\varepsilon_{uv}
    \left\{ \left ( \frac{\sigma_{uv}}{r_{uv}} \right)^{12}
    - \left ( \frac{\sigma_{uv}}{r_{uv}} \right)^{6}
    \right \} \notag \\
    &\quad+q_vV_{\rm ESP}({\bm r}), \label{eq:uvr} 
\end{align}
where $\sigma_{uv}$ and $\varepsilon_{uv}$ are the Lennard-Jones parameters between the solute site $u$ and solvent site $v$,
$q_v$ is the point charge of the solvent site $v$, and $V_{\rm ESP}({\bm r})$ is the electrostatic potential at position ${\bm r}$ depending on the electronic structure of the solute, whose formulation is shown later.

\subsubsection{3D-RISM-SCF theory}

3D-RISM-SCF theory combines electronic structure theory and 3D-RISM.
The procedure involves the self-consistent treatment of solute electrostatic potential and solvent charge density.
Solvent charge density of 3D-RISM is calculated as follows:
\begin{align}
q_I = \sum_{v \in {\rm solvent}} \rho_v q_v g_v({\bm r})\delta^3, \label{eq:solvq}
\end{align}
where $\rho_v$ and $\delta^3$ are the average densities of atom $v$ and the volume element, respectively.

The electrostatic potential of the solute is then given by
\begin{align}
    V_{\rm ESP}({\bm r}_I) &= \sum_{A=1}^{N_{\rm atom}} \frac{Z_A}{|{\bm r}_I-{\bm r}_A|} \notag \\
    &\quad - \sum_{\alpha, \beta} D_{\alpha \beta} \left \langle \chi_\alpha \left | \frac{1}{|{\bm r}_I - {\bm r}'|} \right |\chi_\beta \right \rangle.
\end{align}
The first term is the potential from the nuclei of the solute.
${\bm r}_A$ and $Z_A$ are the coordinate and the nuclear charge of the $A$-th atom. $N_{\rm atom}$ is the number of atoms in the solute.
The second term refers to the electrons, whose coordinate is ${\bm r}'$.
$D_{\alpha \beta}$ is the matrix element of the density matrix, whose indices are the $\alpha$-th and $\beta$-th atomic orbitals.
The electrostatic potential is updated using the density matrix, which is derived from Eq. (\ref{eq:Hsolv}) and delivered to the Eq. (\ref{eq:uvr}) in order to solve the 3D-RISM equation.
Subsequently, the 3D-RISM and closure equations are solved to obtain the solvent charge density [Eq. (\ref{eq:solvq})].

The steps above are repeated until the Helmholtz energy $\mathcal{A}$ converges.
The Helmholtz energy is defined as follows:
\begin{align}
\mathcal{A} &:= E_{\rm solute} + \Delta \mu \\
&\,= \langle \Psi_{\rm solv}|\mathcal{H}_{\rm iso} | \Psi_{\rm solv} \rangle + \Delta \mu.
\end{align}
The first term $E_{\rm solute}$ is the total energy of the solute.
This term is the expectation value of the Hamiltonian of the isolated solute $\mathcal{H}_{\rm iso}$ using the wave function of the solute $\Psi_{\rm solv}$ in a solvent.
The second term is the solvation free energy. The expression of $\Delta \mu$ depends the choice of closure relationship\cite{kovalenko1999self,sato2000self,hirata2003molecular}.

\subsection{3D-RISM-VQE theory} 
We propose a combination of 3D-RISM and the VQE by extending the 3D-RISM-SCF method. 
During VQE, variational optimization is performed to minimize the expectation value $\langle \Psi_{\rm solv}({\bm \theta})|\mathcal{\widetilde{H}}_{\rm solv} | \Psi_{\rm solv}({\bm \theta}) \rangle$ with respect to the rotational angles ${\bm \theta}$ of the quantum circuit.
Hereafter, an expectation value is abbreviated as $\langle \mathcal{\widetilde{H}}_{\rm solv}({\bm \theta}) \rangle$.
The solvated Hamiltonian $\mathcal{H}_{\rm solv}$ must be converted into the Pauli-operators form as follows:
\begin{align}
\mathcal{\widetilde{H}}_{\rm solv} = \sum_{l=1}^L h_l P_l, \label{eq:pauliform}
\end{align}
where $P_l$ is the Pauli operator, and the tilde signifies a qubit representation, and coefficients $\{ h_l \}$ are obtained from molecular integrals.
\begin{figure*}[ht]
 \centering
 \includegraphics[width=18cm, bb=0 0 1324 400]{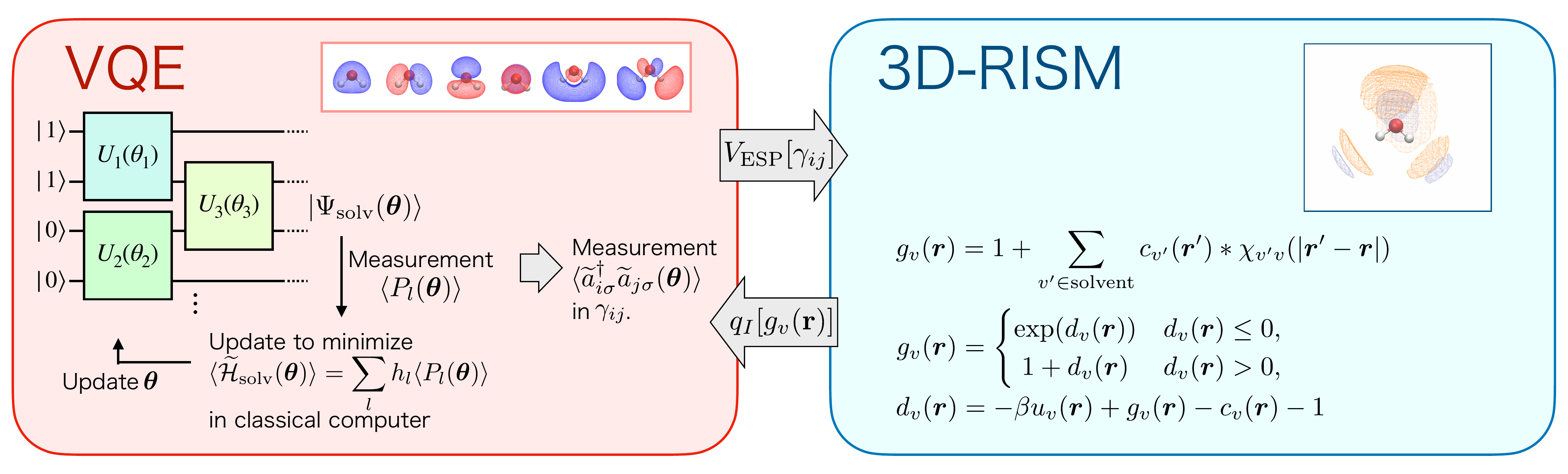}
 \caption{The scheme of 3D-RISM-VQE.}
 \label{fig:sch}
\end{figure*}

VQE and 3D-RISM computations are repeated to obtain the self-consistent solvent charge density and solute density matrix.
In 3D-RISM-VQE, the solute density matrix can be obtained from measurements.
The density matrix ${\bm D}$ is given by the follows:
\begin{align}
    D_{\alpha \beta} = \sum_{i,j} C_{\alpha i}\gamma_{ij}C_{j \beta}^*,
\end{align}
where $C_{\alpha i}$ is the coefficient of atomic orbital $\chi_\alpha$ in the $i$-th molecular orbital.
The one electron reduced density matrix (1-RDM) ${\bm \gamma}$ is represented as follows:
\begin{align}
\gamma_{ij} = 
\begin{cases}
    \sum_{\sigma} \langle \widetilde{a}_{i\sigma}^\dagger \widetilde{a}_{j\sigma} ({\bm \theta})\rangle & \textrm{(active space)} \\
    2\delta_{ij} & \textrm{(core)} \\
    0 & \textrm{(else)},
\end{cases}\label{eq:1rdm}
\end{align}
where $a_{i\sigma}^\dagger$ and $a_{i\sigma}$ are the creation and annihilation operators of the spin orbital $i\sigma$.
The value of ${\bm \gamma}$ in active space can be obtained by measurements of the expectation value 
$\sum_{\sigma} \langle \widetilde{a}_{i\sigma}^\dagger \widetilde{a}_{j\sigma} ({\bm \theta})\rangle$ via a qubit representation using a quantum computer.

We summarize to show the self-consistent scheme of the 3D-RISM-VQE in Figure \ref{fig:sch}.
The expectation value $\langle \mathcal{\widetilde{H}}_{\rm solv} ({\bm \theta})\rangle$ is minimized with respect to the rotational angle ${\bm \theta}$ of the quantum circuit.
Using the variationally optimized wave function, the value of 1-RDM in Eq. (\ref{eq:1rdm}) is computed, and $V_{\rm ESP}({\bm r}_I)$ is prepared for 3D-RISM calculations.
To prepare $\mathcal{H}_{\rm solv}$, the 3D-RISM equation (Eq. (\ref{eq:RISM})) and a closure relation are solved to obtain $g_v({\bm r})$ and $\Delta \mu$.
$\mathcal{H}_{\rm solv}$ is updated using the new solvent charge density $q_I$.
Finally, the Helmholtz energy of the 3D-RISM-VQE can be obtained using the self-consistent procedure between VQE and 3D-RISM calculations.

\subsection{$L^1$-norm from the Pauli-operator formed Hamiltonian}

The $L^1$-norm of the $L$-dimensional vector ${\bm h} = [h_1\, h_2\, \cdots \,h_L] \in \mathbb{R}^L$, which is a coefficient vector of the qubit-represented Hamiltonian such as Eq. (\ref{eq:pauliform}), is defined as follows
\cite{Berry2020timedependent}:
\begin{align}
\lambda := \| {\bm h} \|_1 = \sum_{l=1}^L |h_l|.
\end{align}
In this paper, $L^1$-norm values do not include the coefficient of the identity operator $I$.

\section{\label{sec:comput} Computational Details}
We have implemented the 3D-RISM-VQE method 
using PySCF\cite{sun2020recent,sun2018pyscf}, OpenFermion\cite{McClean2020}, and Qulacs\cite{suzuki2021qulacs} libraries.
PySCF was used to calculate molecular integrals, natural orbitals, and second-quantized electronic Hamiltonians.
OpenFermion was used to transform a fermionic operator into a qubit representation.
We employed the Jordan--Wigner (JW) transformation for fermion-to-qubit mapping.
Qulacs was employed as a quantum circuit emulator.
We utilized the disentangled \textcolor{black}{unitary coupled-cluster singles and doubles (}UCCSD) \textit{ansatz}\cite{kutzelniggQuantumChemistryFock1982,kutzelniggQuantumChemistryFock1983,kutzelniggQuantumChemistryFock1985,bartlettAlternativeCoupledclusterAnsatze1989,kutzelniggErrorAnalysisImprovements1991,taubeNewPerspectivesUnitary2006,peruzzo2014variational,evangelista2019exact} for VQE.
\textcolor{black}{The active space of VQE is denoted as ($m$e, $n$o), where $m$ and $n$ are the number of electrons and spatial orbitals, respectively.}

3D-RISM calculations were performed using the RISMiCal program package\cite{yoshida2020reference}.
3D-RISM uses Kovalenko--Hiraka (KH) closure\cite{kovalenko1999self,sato2000self}, and water as the solvent.
The number of reference points for the 3D-RISM calculation is given as a three-dimensional grid. For each direction of the $x$, $y$, and $z$-axis, $128$ points were generated with $\delta = 0.25$ \AA \,spacing.
TIP3P\cite{jorgensen1983comparison} with modified hydrogen parameters and OPLS-AA\cite{jorgensen1986monte} models was used for H$_2$O and NH$_4^+$, respectively, regarding the Lennard-Jones parameters for the solutes. The modified hydrogen parameters were $\sigma_H=0.4$ \AA and $\varepsilon_H=192.5$ J mol$^{-1}$.

The geometry of the solutes was optimized using the Gaussian16 package Revision C\cite{g16}. 
Geometry optimizations were performed at the B3LYP/6-31G* level using the PCM.
For Edminston--Ruedenberg (ER) localization in $L^1$-norm analysis, we use a fortran code \texttt{fcidump\_rotation.f90}, which is part of the NECI program package\cite{guther2020neci}.
We performed the ER localization with all occupied and virtual orbitals, and a convergence criterion was $1.0 \times 10^{-6}$.

Three-dimensional \textcolor{black}{molecular} figures of the distribution functions of the solvent and molecular orbitals were visualized using VMD\cite{humphrey1996vmd}.

\section{\label{sec:numeric} Result and discussion}

First, we show the distribution functions of solvent \textcolor{black}{water} $g_{v}({\bm r})$, which is self-consistently determined for the electronic ground-state of the solute \textcolor{black}{water molecule}.
Second, we discuss the potential and Helmholtz energy curves of a NaCl molecule in water.
Next, the Helmholtz energy and the related quantities of H$_2$O and NH$_4^+$ in water were investigated.
Finally, the $L^1$-norms of the solvated Hamiltonian in a qubit representation are shown to elucidate the degree to which the solvent effect influences the difficulty of quantum computations.

\subsection{Spatial distribution of solvent}

\begin{figure}[h]
 \centering
 \includegraphics[width=6cm, bb=0 0 1700 1522]{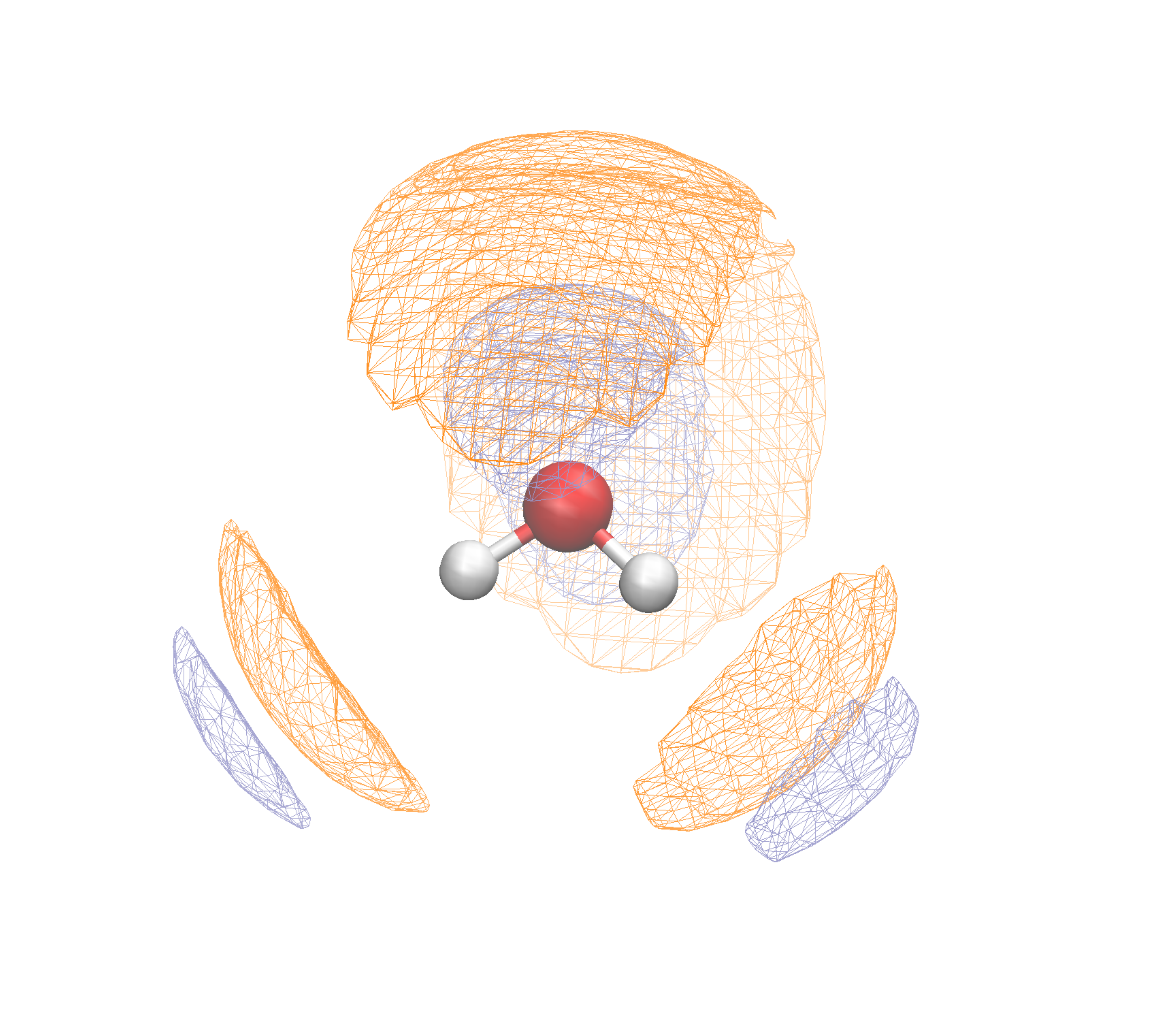}
 \caption{Spatial distribution functions of water around H$_2$O. Orange and ice-blue colorations correspond to $g_{{\rm O}}({\bm r})$ and $g_{{\rm H}}({\bm r})$. Isovalues are 2.4 and 1.8, respectively.}
 \label{fig:sdf}
\end{figure}

To determine the solvation structure intuitively, we first visualized its three-dimensional distribution functions $g_{v}({\bm r})$ in Figure \ref{fig:sdf}, resulting from the 3D-RISM-VQE(8e, 6o) calculation with the 6-31G* basis set.
The arched bilayered solvent shell near the solute oxygen and on the opposite side of the two hydrogens of the solute can be identified; the inside layer is $g_{{\rm H}}({\bm r})$ and the outside is $g_{{\rm O}}({\bm r})$.
Two other distributions of $g_{{\rm O}}({\bm r})$ and $g_{{\rm H}}({\bm r})$ are situated around the two hydrogen atoms and on the opposite side of the solute oxygen.

\begin{figure}[htbp]
  \begin{minipage}{1\linewidth}
    \centering
    \includegraphics[width=8.5cm,bb=156 473 450 685]{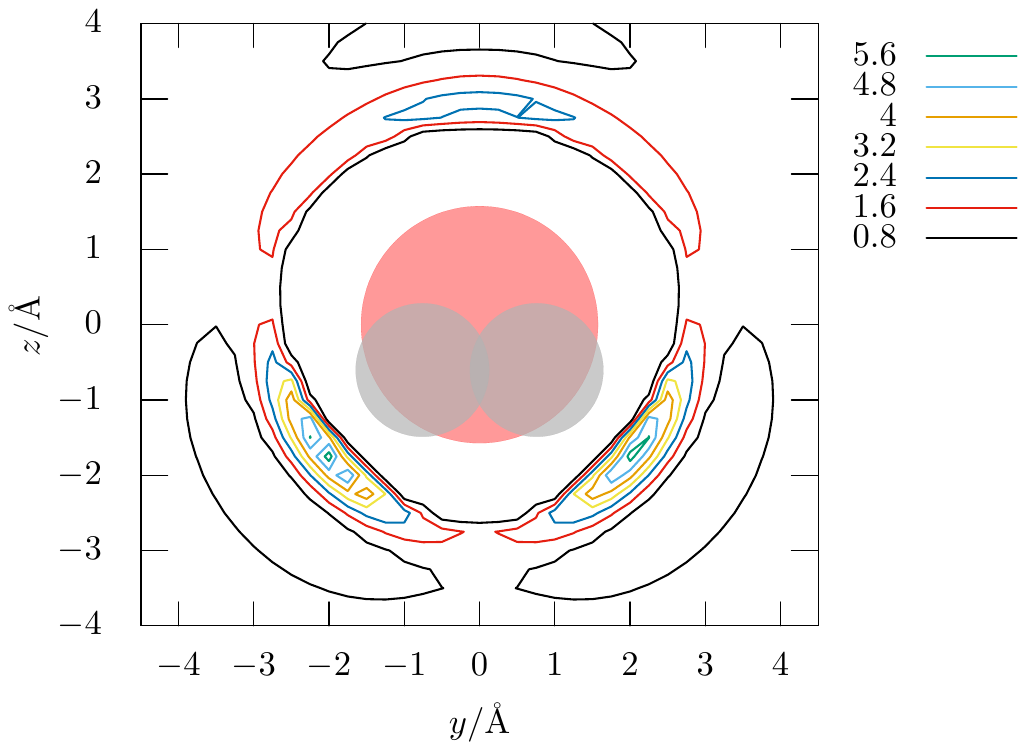}
    \subcaption{$g_{{\rm O}}({\bm r})$}
  \end{minipage}
  \begin{minipage}{1\linewidth}
    \centering
    \includegraphics[width=8.5cm,bb=156 473 450 685]{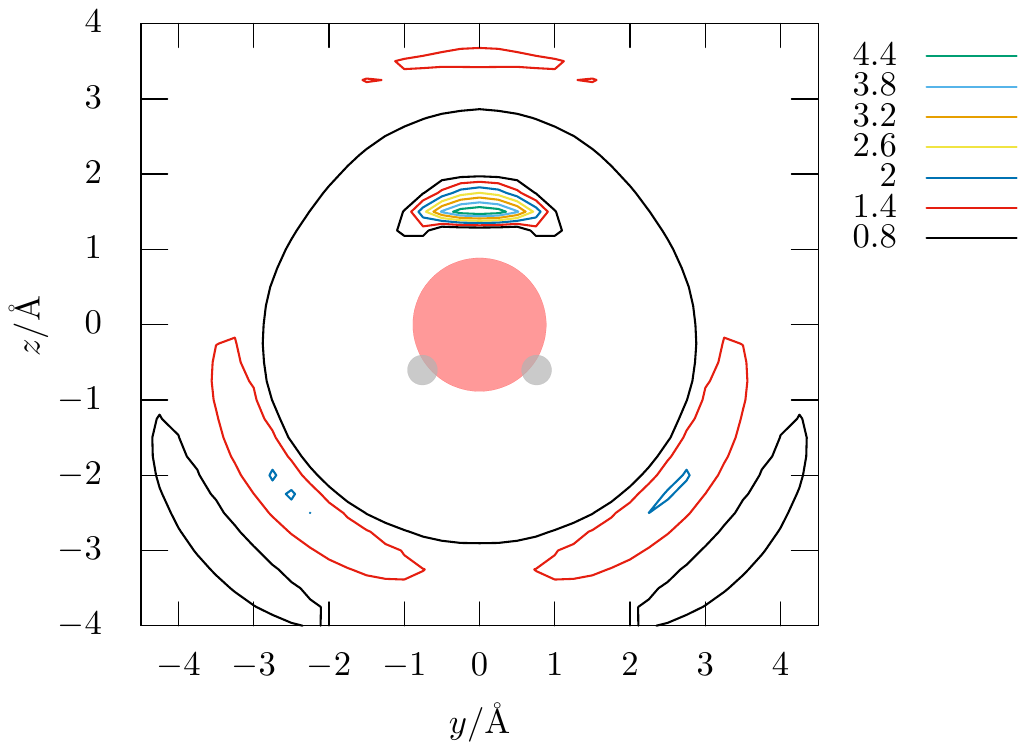}
    \subcaption{$g_{{\rm H}}({\bm r})$}
  \end{minipage}
  \caption{Contour plot showing the spatial distributions of water around an H$_2$O molecule on the $x=0$ plane. The diameter of the solute atoms is $\sigma_{uv}$. \label{fig:guv}}
\end{figure}

Contour plots of the distribution functions $g_{v}({\bm r})$ are shown in Figure \ref{fig:guv}.
The solute water molecule is positioned on the $x=0$ plane, and the distribution functions on the same plane are shown.
The resulting distribution functions $g_{v}({\bm r})$ are reliable compared to previous 3D-RISM studies\cite{beglov1997integral}.
Figure \ref{fig:guv} (a) shows the distribution when $v={\rm O}$ and the two significant peaks exist, symmetrically, close to the solute H and on the other side of the solute O across from the solute H.
These peaks occur because solute H atoms are polarized \textcolor{black}{positively} and attract the \textcolor{black}{negatively} charged O site via electrostatic interaction. 
A small peak around $(y,z)=(0,3)$ 
is related to the distribution of solvent H atoms.

Figure \ref{fig:guv} (b) shows that the distribution $g_{{\rm H}}({\bm r})$ indicating a sharp peak around $(y,z)=(0,1.5)$.
This peak indicates that the \textcolor{black}{negatively} polarized solute O atom is attracted to the solvent H sites, which are \textcolor{black}{positively} charged.
The small peak of $g_{{\rm O}}({\bm r})$ around $(y,z)=(0,3)$ is related to this prominent peak because of intramolecular correlation.
Similarly, the two small peaks of $g_{{\rm H}}({\bm r})$ in the lower part of the figure are correlated to the two large peaks of $g_{{\rm O}}({\bm r})$.
Therefore, 3D-RISM-VQE can calculate appropriate distribution functions of solvent water.

\subsection{Energy curves of NaCl}

\begin{figure}[ht]
 \centering
 \includegraphics[width=8cm, bb=0 0 460 474]{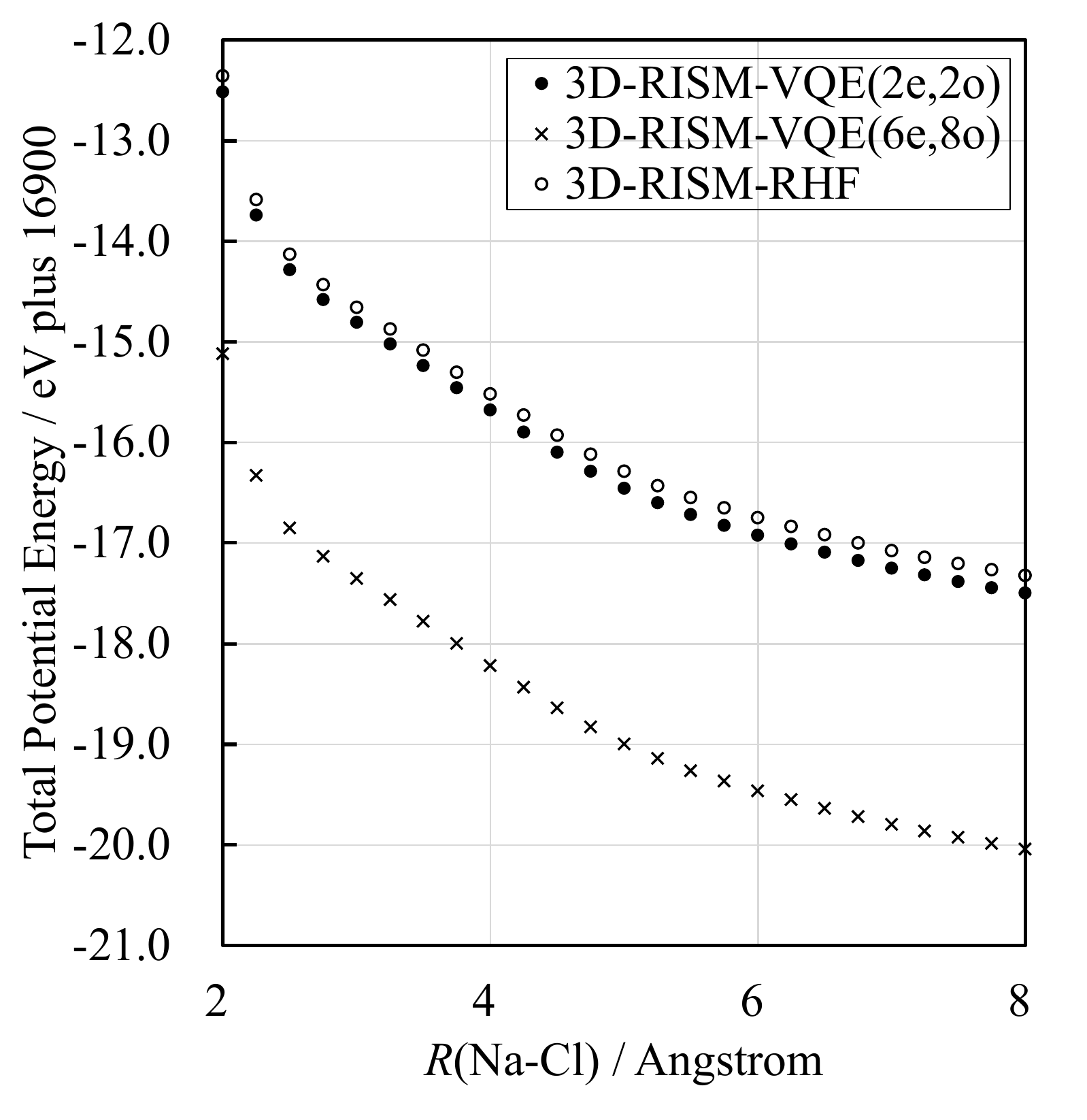}
 \caption{Potential energy curve of NaCl calculated by combined 3D-RISM and VQE or restricted Hartree-Fock (RHF), respectively.} 
 \label{fig:pec1_pot}
\end{figure}

Figure \ref{fig:pec1_pot} shows the potential energy curve of NaCl dissociation\textcolor{black}{, where the points were calculated at 0.25 \AA \,intervals.}
The cc-pVTZ basis set was chosen.
The potential energy is defined as the sum of the expectation value of $\mathcal{H}_{\rm iso}$ and the solute--solvent binding energy\textcolor{black}{\cite{okamoto2019implementation}}.
We employed the natural orbitals \textcolor{black}{of CCSD} to include electron correlation effectively.
\textcolor{black}{The potential energies of 3D-RISM-VQE(2e, 2o) stabilize at approximately 0.15 eV at 3.00 \AA\, and 0.18 eV at 7.00 \AA\, compared with the 3D-RISM-SCF calculations using restricted Hartree-Fock (RHF), termed 3D-RISM-RHF}.
\textcolor{black}{The energy difference between them is slightly larger in the region where the interatomic distance $R$ is long.}
The curve of 3D-RISM-VQE(6e, 8o) becomes more stable at approximately 2.7 eV as a result of the electron correlation effect.
\textcolor{black}{The largest stabilization is 2.76 eV at 2.00 \AA.}

The potential energy curve does not exhibit any minima in this region.
In the gas phase, the two ionic species may have made contact \textit{via} electrostatic interactions\cite{zeiri1983theory}.
On the other hand, in water, this figure shows the dissociative features of NaCl without forming a chemical bond \textcolor{black}{from the viewpoint of total potential energy}.
This agrees with the conclusions of the one-dimensional RISM-MCSCF calculations\cite{okamoto2019implementation}.

\begin{figure}[ht]
 \centering
 \includegraphics[width=8cm, bb=0 0 460 474]{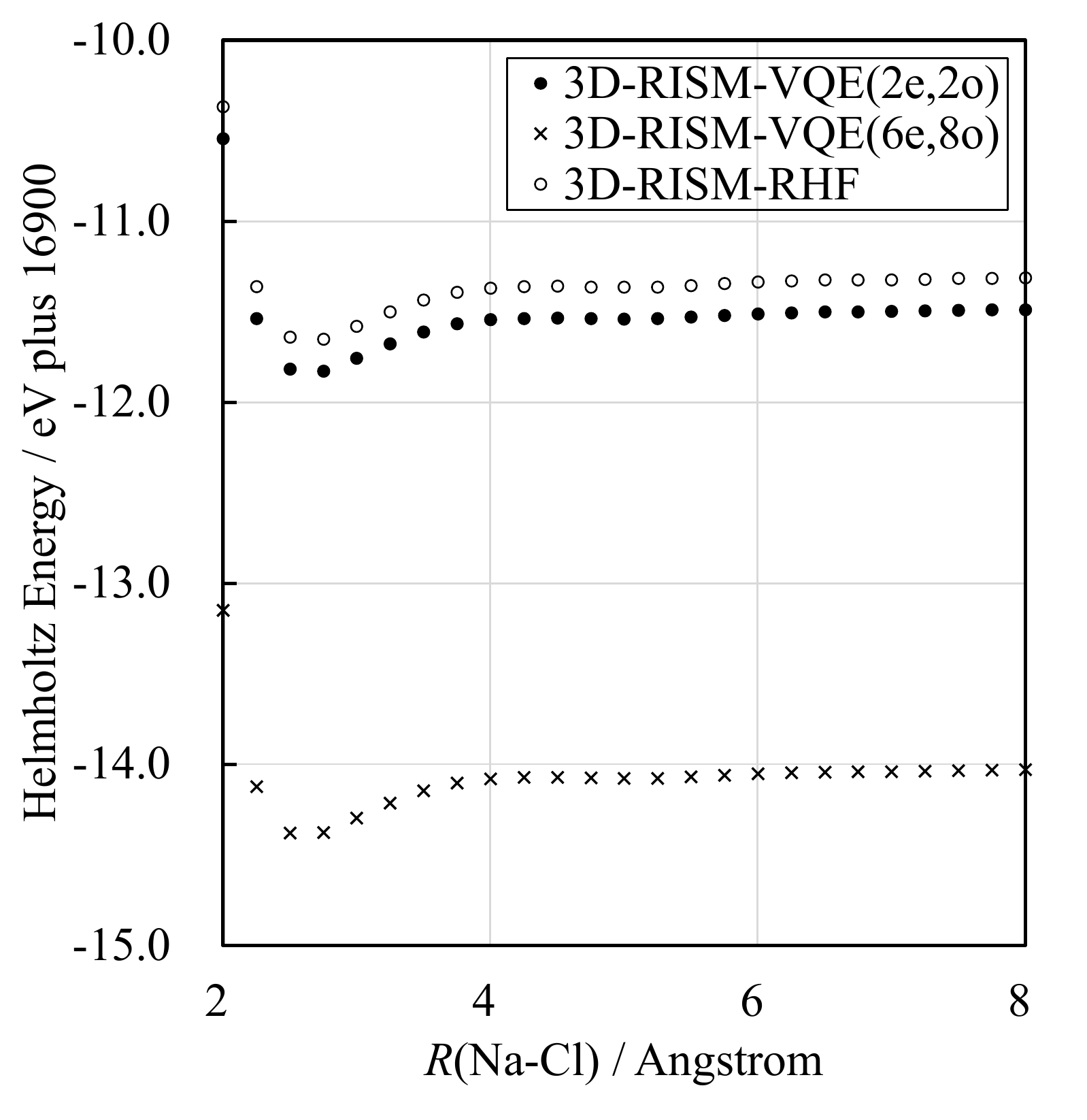}
 \caption{The Helmholtz energy curve of NaCl calculated by combined 3D-RISM and VQE or RHF, respectively.} 
 \label{fig:pec1_free}
\end{figure}

Figure \ref{fig:pec1_free} shows the Helmholtz energy curve of NaCl.
Unlike the potential energy, an energy minimum around \textcolor{black}{$R/{\rm \AA} =2.50$ to $2.75$} is present.
At this bond length, the ions are considered to be in contact.
\textcolor{black}{In the (2e, 2o) case, the Helmholtz energy difference from 3D-RISM-RHF is nearly constant, about $-0.17$ to $-0.18$ eV.}
\textcolor{black}{In the (6e, 8o) case, the difference is also nearly constant in the R/\AA $\geq$ 2.75 region, about $-2.71$ to $-2.72$ eV.
The difference in the tightly contact ion pair region becomes slightly larger, and the maximum is $-2.78$ eV at 2.00 \AA.}
\textcolor{black}{Stabilizations in the two VQEs are almost independent of the interatomic distance because the natural orbitals in the active space are localized in the Cl atom.}

\textcolor{black}{Another shallow minimum is found at around 5.00 \AA \,in the two 3D-RISM-VQE and -RHF calculations.}
\textcolor{black}{At this minimum, the two ions are not in contact, but the solvent waters are distributed between them.}
\textcolor{black}{This is a feature of the so-called solvent-shared ion pair\cite{hirata2003molecular,Brini2017how,yao2018free}.}

The Helmholtz energy of the ionic pair dissociation determined by our calculations is not lower than that of the minimum of the contact ions; this agrees with the results of an \textit{ab initio} molecular dynamics study\cite{yao2018free}, although the statistical ensemble is not the same as RISM.

\subsection{H$_2$O and NH$_4^+$}

\subsubsection{Helmholtz energies}

Table \ref{tab:AHsolv} shows the values of the Helmholtz energy $\mathcal{A}$, the free energy components $\langle \mathcal{H_{\rm iso}} ({\bm \theta})\rangle$ and $\Delta \mu$, and the expectation value of the solvated Hamiltonian $\langle \mathcal{H_{\rm solv}} ({\bm \theta})\rangle$.
This table shows that the solute energy $\langle \mathcal{H_{\rm iso}} ({\bm \theta})\rangle$ predominantly contributes to the free energy $\mathcal{A}$.
The value of $\mathcal{A}$ decreases systematically in both molecules as the basis set or active space becomes larger.
The energy of the solute $\langle \mathcal{H_{\rm iso}} ({\bm \theta})\rangle$ has the same trend as $\mathcal{A}$, and the contribution of $\Delta \mu $ to the free energy is minimal.

As the active space expands, the value of the solvation free energy $\Delta \mu$ increases less than the decrease of $\langle \mathcal{H_{\rm iso}} ({\bm \theta})\rangle$.
It is considered that the electronic interaction between solute and solvent is given by the multiplication of the distributions of the solvent and electrons \textcolor{black}{and} a small part of the electrons move into virtual orbitals to stretch the distribution \textit{via} electronic correlation.
However, it should be reiterated that the contribution of $\Delta \mu$ is much smaller than that of $\langle \mathcal{H_{\rm iso}} ({\bm \theta})\rangle$.

The value of $\langle \mathcal{H_{\rm solv}} ({\bm \theta})\rangle$ is comparable to that of $\langle \mathcal{H_{\rm iso}} ({\bm \theta})\rangle$, and the contribution of solute--solvent interaction $\langle \mathcal{V_{\rm solv}} ({\bm \theta})\rangle$ \textcolor{black}{is} minor in comparison.
Regarding $\langle \mathcal{H_{\rm solv}} ({\bm \theta})\rangle$, there was no clear increasing or decreasing trend against the sizes of the basis set and the active space.

\begin{table*}[hbtp]
    \caption{Helmholtz energy components and expectation values of solvated Hamiltonians of H$_2$O and NH$_4^+$ in aqueous solution (in a.u.).}
    \label{tab:AHsolv}
    \centering
    \begin{tabular}{l l l r r r r} \hline
        Molecule & Basis set & Method & \multicolumn{1}{c}{$\mathcal{A}$} & \multicolumn{1}{c}{$\langle \mathcal{H}_{\rm iso} ({\bm \theta})\rangle$} & \multicolumn{1}{c}{$\Delta \mu$} & \multicolumn{1}{c}{$\langle \mathcal{H}_{\rm solv} ({\bm \theta})\rangle$} \\ \hline \hline
        H$_2$O & STO-3G & 3D-RISM-VQE(2e, 2o) & $-$74.9651 & $-$74.9651 &  0.0001 & $-$75.3240 \\
               &        & 3D-RISM-VQE(8e, 6o) & $-$75.0138	& $-$75.0152 &  0.0014 & $-$75.3489 \\ \cline{2-7}
               & 6-31G  & 3D-RISM-VQE(2e, 2o) & $-$75.9988 & $-$75.9781 & $-$0.0207 & $-$76.6014 \\
               &        & 3D-RISM-VQE(8e, 6o) & $-$76.0086 & $-$75.9905 & $-$0.0181 & $-$76.5833 \\\cline{2-7}
               & 6-31G* & 3D-RISM-VQE(2e, 2o) & $-$76.0178 & $-$76.0044 & $-$0.0134 & $-$76.5387 \\
               &        & 3D-RISM-VQE(8e, 6o) & $-$76.0271 & $-$76.0159 & $-$0.0112 & $-$76.5225 \\ \hline \hline
        NH$_4^+$ & STO-3G & 3D-RISM-VQE(2e, 2o) & $-$55.9672 & $-$55.8685 & $-$0.0987 & $-$53.6213 \\
                 &        & 3D-RISM-VQE(4e, 4o) & $-$55.9693 & $-$55.8706 & $-$0.0987 & $-$53.6235 \\
                 &        & 3D-RISM-VQE(6e, 6o) & $-$55.9977 & $-$55.8990 & $-$0.0986 & $-$53.6509 \\
                 &        & 3D-RISM-VQE(8e, 8o) & $-$56.0462 & $-$55.9478 & $-$0.0983 & $-$53.7024 \\ \cline{2-7}
                 & 6-31G  & 3D-RISM-VQE(2e, 2o) & $-$56.6154 & $-$56.5161 & $-$0.0993 & $-$54.2678 \\
                 &        & 3D-RISM-VQE(4e, 4o) & $-$56.6159 & $-$56.5166 & $-$0.0993 & $-$54.2670 \\
                 &        & 3D-RISM-VQE(6e, 6o) & $-$56.6235 & $-$56.5244 & $-$0.0991 & $-$54.2764 \\
                 &        & 3D-RISM-VQE(8e, 8o) & $-$56.6365 & $-$56.5376 & $-$0.0989 & $-$54.2906 \\ \cline{2-7}
                 & 6-31G* & 3D-RISM-VQE(2e, 2o) & $-$56.6287 & $-$56.5294 & $-$0.0994 & $-$54.2806 \\
                 &        & 3D-RISM-VQE(4e, 4o) & $-$56.6293 & $-$56.5299 & $-$0.0994 & $-$54.2798 \\
                 &        & 3D-RISM-VQE(6e, 6o) & $-$56.6366 & $-$56.5373 & $-$0.0992 & $-$54.2882 \\
                 &        & 3D-RISM-VQE(8e, 8o) & $-$56.6492 & $-$56.5502 & $-$0.0990 & $-$54.3030 \\ \hline
    \end{tabular}
\end{table*}

\subsubsection{Evaluation of 1-norm}

The $L^1$-norm values of the Hamiltonians with and without solvent effect are compared in Table \ref{tab:lmdb}.
\textcolor{black}{The values of $\lambda$ by Bravyi--Kitaev (BK) and symmetry conserving BK transformations are the same as those obtained by JW transformation.}
Table \ref{tab:lmdb} shows that the $\lambda$ value using canonical molecular orbital (CMO) is almost unchanged when adding the solvent effect; the ER-localized orbital indicates a similar trend.
This can be explained if the expectation values of $\mathcal{H_{\rm solv}}$ and $\mathcal{H_{\rm iso}}$ are comparable, as shown in Table \ref{tab:AHsolv}, and implies that a quantum computer could perform quantum chemical calculations of a molecule in solution at practically the same cost as in gas.

Koridon et al. demonstrated that orbital localizations could decrease the $L^1$-norm value compared to CMO\cite{koridon2021orbital}.
However, the reduction effect is not apparent in our analysis. Our molecules are small, and the difference between delocalized and localized orbitals is also small.


\begin{table*}[hbtp] 
  \caption{Values of $\lambda$ with the methods for molecules in gas (g) and in solution (s).} 
  \label{tab:lmdb}
  \centering
    \begin{tabular}{llrrrrrrrrrrrr}
    \hline
    Molecule & Method & STO-3G &&&& 6-31G &&&& 6-31G* &&& \\ 
    && CMO & s/g ratio & ER & s/g ratio & CMO & s/g ratio & ER & s/g ratio & CMO & s/g ratio & ER & s/g ratio \\ \hline \hline
    H$_2$O & RHF & 72 & & 80 && 160 && 139 && 327 && 365 & \\
     & 3D-RISM-VQE(2e, 2o) & 72 & 100.3\% & 81 & 100.3\% & 153 & 96.0\% & 139 & 99.5\% & 326 & 99.8\% & 364 & 99.8\% \\
     & 3D-RISM-VQE(8e, 6o) & 72 & 100.3\% & 81 & 100.3\% & 153 & 96.1\% & 139 & 100.0\% & 328 & 100.3\% & 364 & 99.9\% \\ \hline
    NH$_4^+$ & RHF & 73 & & 78 && 230 && 178 && 475 && 429 & \\
     & 3D-RISM-VQE(2e, 2o) & 71 & 97.2\% & 76 & 97.4\% & 233 & 101.5\% & 181 & 101.9\% & 480 & 101.0\% & 434 & 101.1\% \\
     & 3D-RISM-VQE(4e, 4o) & 71 & 97.3\% & 76 & 97.4\% & 233 & 101.6\% & 181 & 101.8\% & 480 & 101.0\% & 434 & 101.1\% \\
     & 3D-RISM-VQE(6e, 6o) & 71 & 97.4\% & 76 & 97.4\% & 234 & 101.9\% & 180 & 101.5\% & 481 & 101.1\% & 434 & 101.2\% \\
     & 3D-RISM-VQE(8e, 8o) & 71 & 97.2\% & 76 & 97.4\% & 233 & 101.5\% & 181 & 101.9\% & 480 & 101.0\% & 434 & 101.1\% \\ \hline
    \end{tabular} 
\end{table*}


\section{\label{sec:conclusion} Conclusion}

3D-RISM-VQE combines VQE and 3D-RISM-SCF theory.
The electronic structure of the solute is affected by the countless number of solvent molecules, and the electrostatic interaction term from the solvent is added into the solute Hamiltonian in a QM/MM manner.
Using an analytical treatment of the solvent distribution, the present method does not include errors from the solvent configuration sampling.

We have demonstrated the spatial distribution functions $g_{v}({\bm r})$ around an H$_2$O, showing that the present method can give reasonable distributions of solvent water.
We have also shown the potential and Helmholtz energy curves of NaCl.
The dissociative feature of the potential energy curve, previously shown by one-dimensional RISM-MCSCF\textcolor{black}{\cite{okamoto2019implementation}}, was successfully reproduced.
The minimum of the Helmholtz energy curve corresponds to the ions in contact.
A dissociation of molecular species is one of the essential themes that can cause static electronic correlation.
In the case of H$_2$O and NH$_4^+$, the Helmholtz energy components and the expectation values of the solvated Hamiltonians were investigated. The expectation values of the solute Hamiltonian are dominant in both the Helmholtz energy and the expectation value of the solvated Hamiltonian.
Finally, we analyzed the $L^1$-norm of the modified Hamiltonian in a qubit representation, which is a key to quantifying the algorithmic scaling of various quantum algorithms.
The $L^1$-norm value of the solvated Hamiltonian is almost the same as that in the gas phase, implying that the efficiency of quantum chemical computations including solvent effects on a quantum computer are virtually the same as those in gas phases.

\begin{acknowledgments}
This work was supported by MEXT Quantum Leap Flagship Program (MEXT Q-LEAP) Grant Number JPMXS0118067394 and JPMXS0120319794.
W.M. wishes to thank Japan Society for the Promotion of Science (JSPS) KAKENHI No.\ \textcolor{black}{JP}18K14181 and JST PRESTO No.\ JPMJPR191A.
Y.Y. wishes to thank the financial support of JSPS KAKENHI Grant Number JP21K20536.
W.M. and Y.Y. also acknowledge support from JST COI-NEXT program Grant No. JPMJPF2014. 
N.Y. is grateful to the JSPS KAKENHI No.\ \textcolor{black}{JP}19H02677 and \textcolor{black}{JP}22H05089.
The computations were performed using the Institute of Solid State Physics at the University of Tokyo, the Research Institute for Information Technology (RIIT) at Kyushu University, Japan, the SQUID supercomputer at the Cybermedia Center of Osaka University, and Research Center for Computational Science, Okazaki, Japan (Project: 21-IMS-C244, 22-IMS-C076, 22-IMS-C176).
\end{acknowledgments}


\bibliography{main}

\end{document}